\documentclass[useAMS,usenatbib]{mn2e}
\usepackage{graphicx}
\usepackage{amssymb}
\usepackage{amsmath}
\usepackage[usenames]{color}
\definecolor{DarkBlue}{rgb}{0.1,0,0.6}

\newcommand{\myr}{\mathrm{Myr}}

\title[Evolution of stellar disc in Galactic Centre]{Secular theory of
the orbital evolution of the young stellar disc in the Galactic Centre}
\author[J. Haas, L. \v{S}ubr and
  D. Vokrouhlick\'{y}]{J. Haas$^{1}$\thanks{E-mail:
  haas@sirrah.troja.mff.cuni.cz},
  L. \v{S}ubr$^{1,2}$ and D. Vokrouhlick\'{y}$^{1}$\\
  $^{1}$Astronomical Institute, Faculty of Mathematics and Physics,
  Charles University, V Hole\v{s}ovi\v{c}k\'{a}ch
  2, 18000 Praha, Czech Republic\\
  $^{2}$Astronomical Institute, Academy of Sciences, Bo\v{c}n\'{i} II,
  14131 Praha, Czech Republic}
\begin{document}
\date{Accepted ---. Received ---; in original form ---}
\pagerange{\pageref{firstpage}--\pageref{lastpage}} \pubyear{2011}
\maketitle
\label{firstpage}
%
%
\begin{abstract}
We investigate
the orbital evolution of a system of $N$ mutually
interacting stars on initially circular orbits around the dominating central
mass.
We include perturbative
influence of a distant axisymmetric source and an extended
spherical potential. In particular, we focus on the case when the secular
evolution of orbital eccentricities is suppressed by the spherical perturbation.
By means of standard perturbation methods, we derive
semi-analytic formulae for the evolution of normal vectors of the individual
orbits. We find its two qualitatively different modes. Either the orbits
interact strongly and, under such circumstances, they
become dynamically coupled, precessing synchronously in the potential of the
axisymmetric perturbation. Or, if their mutual interaction is weaker, the orbits
precess
independently, interchanging periodically their angular momentum, which leads
to oscillations of inclinations. We argue that these processes may have been
fundamental for the evolution of the disc of young stars orbiting the
supermassive black hole in the centre of the Milky Way.
\end{abstract}
%
%
\section{Introduction}
Problem of dynamics in the perturbed Keplerian potential has been studied
extensively throughout the history of celestial mechanics. Due to high
attainable
accuracy of observational data, its primary field of application has always
been the Solar System, which naturally influenced the selection
of included perturbations. Ones of those widely considered are, due to their
resemblance with the averaged motion of planets, axisymmetric
gravitational potentials.

The above problem has, however, also been investigated for systems
with larger length scales, such as dense star clusters. In that case,
the source of the Keplerian
potential is often
represented by a supermassive black hole (SMBH) which is widely assumed to
reside
in the centres of such clusters. Axisymmetric perturbation is then either due
to a secondary massive black hole (e.g. Ivanov, Polnarev \& Saha 2005)
or a gaseous disc or
torus (e.g. Karas \& \v{S}ubr 2007).
It turns out that in these systems, the
secular evolution of individual stellar orbits is, beside the axisymmetric
perturbation, also affected by a possible
additional spherical potential. Such a potential
may be generated by a stellar cusp or it can represent a post-Newtonian
correction
to the gravity of the central black hole.

In this paper, we extend the analyses of previous authors by means
of standard
tools of celestial mechanics. Our main aim is to incorporate mutual interaction
of stars on nearly-circular orbits around the dominating central mass whose
potential
is perturbed by a distant axisymmetric source and an extended spherical
potential. We apply our results to the observed system of young stars
\citep{Genzel03, Ghez05, Paumard06, Bartko09, Bartko10} orbiting
the SMBH of mass $M_{\bullet}\approx4\times10^6\,M_{\odot}$ \citep{Ghez03,
Eisenhauer05, Gillessen09a, Gillessen09b, Yelda10}
in the centre of the Milky Way.
As an axisymmetric perturbation to its gravity we consider a massive molecular
torus (the so-called circumnuclear disc; CND) which is located at radius
$R_{\mathrm{CND}}\approx1.8~\mathrm{pc}$ from the centre \citep{Christopher05}.
Finally, we
consider gravity
of a roughly spherical cusp of late-type stars \citep{Genzel03, Schoedel07,
Dot09} which is believed to be present in this region, as well.
Within this
context, we broaden the analysis of
our previous paper (Haas, \v{S}ubr \& Kroupa 2011) where we have studied
the dynamical evolution of this kind of system purely by means of numerical
$N$-body calculations. In particular, we now develop a simple semi-analytic
model which naturally explains key features of our prior results.

The paper is organized as follows.
In the theoretical Section~\ref{sec:theory},
we first discuss the influence of the spherical perturbative potential upon the
stellar
orbits (Section~\ref{eccom}). This allows us to separate the evolution of
eccentricity
from the rest of
the problem and, subsequently,
to formulate equations for the evolution of
inclinations and
nodal longitudes (Section~\ref{incom}).
In Section~\ref{sec:sgra}, we present an example
of the orbital evolution
of a stellar disc motivated by the configuration that is observed in the
Galactic Centre. We conclude our results in Section~\ref{sec:conclusions}.
%
%
\section{Theory}
\label{sec:theory}
To set the stage, we first develop a secular theory of orbital evolution
for two (later in the section generalized to multiple) stars orbiting a
massive centre, the SMBH, taking into account their mutual
gravitational interaction and perturbations from the spherical stellar
cusp and the axisymmetric CND.
The CND is considered stationary and
its model is further simplified and
taken equivalent to a ring at a certain distance from the centre.
It should be, however, pointed out that generalization to a more realistic
structure, such as thin or thick disc, is straightforward in our setting
but we believe at this stage it would just involve algebraic complexity
without bringing any new quality to the model. In the same way,
the stellar cusp
is reduced to an equilibrium spherical model without
involving generalizations beyond that level. For instance, an axisymmetric
component of the stellar cusp may be effectively accounted for by the
CND effects in the first approximation.

We are going to use standard tools of classical celestial mechanics,
based on the first-order secular solution using the perturbation methods
(see, e.g., Morbidelli 2002 or Bertotti, Farinella \& Vokrouhlick\'{y}
2003 for general discussion).
In particular, the stellar orbits are described using a conventional set of
Kepler's elements which are assumed to change according to Lagrange
equations. Since we are interrested in a long-term
dynamical evolution of the stellar orbits we replace the perturbing
potential (or potential energy) with its average value over one revolution
of the stars about the centre, which is the proper sense of addressing
our approach as secular. In doing so, we assume there is no orbital
mean motion resonance between the two (or multiple) stars. As an implication
of our approach, the orbital semi-major axes of the stellar orbits are
constant and information about the position of the stars in orbit is irrelevant.
The secular system thus consists of description how the remaining four
orbital elements, eccentricity, inclination, longitude of node and
argument of pericentre, evolve in time. This is still a very complicated
problem in principle, and we shall adopt simplifying assumption that
will allow us to treat the eccentricities and pericentres separately
(Section~\ref{eccom}) and leave us finally with the problem of dynamical
evolution of inclinations and nodes (Section~\ref{incom}). Note this is
where our approach diverges from typical applications in planetary
systems, in which this separation is often impossible.
%
%
\subsection{Confinement of eccentricity}\label{eccom}
In this section we discuss our assumptions about eccentricity and pericentre
evolution. For this moment, we drop the mutual interaction of stars from
our consideration. We assume that the initial stellar orbits have small
eccentricity and we describe under which conditions we may assume they stay
small to the point we could neglect them. Note this is not an obvious
conclusion because axially symmetric systems (such as a perturbing massive
ring) have been extensively studied in planetary applications and it has been
shown that non-conservation of the total orbital angular momentum may
lead to a large, correlated variations of eccentricity and inclination
even if the initial eccentricity is arbitrarily small. This is often called
Kozai secular resonance as a tribute to a pioneering work of Kozai (1962)
(see also Lidov 1962).
In what follows we describe conditions under which this process is
inhibited in our model.
%
%
\subsubsection{Stellar cusp potential}\label{sc}
We start with our assumption about the potential energy of a star of mass
$m$ in the spherical cusp of the
late-type stars surrounding the centre. Considering a general power-law radial
density profile of the cusp, $\rho\left(r\right)\propto r^{-\alpha}$, we have
the potential energy
\begin{eqnarray}
  {\cal R}_{\mathrm{c}} = -\frac{GmM_{\mathrm{c}}}{\beta R_{\mathrm{CND}}}
  \left(\frac{r}{R_{\mathrm{CND}}}\right)^{\beta},&&
\label{bw2}
\end{eqnarray}
where $\beta=2-\alpha$, the cusp mass within a scale
distance $R_{\mathrm{CND}}$ is denoted $M_{\mathrm{c}}$ and
$G$ stands for the gravitational constant.
According to the averaging technique, we shall integrate the potential
energy (\ref{bw2}) over one revolution about the centre with respect to the
mean anomaly $l$,
\begin{eqnarray}
  {\overline{\cal R}}_{\mathrm{c}}\equiv\frac{1}{2\pi}
  \int\limits_{-\pi}^{\pi}\mathrm{d}l\;\,{\cal R}_{\mathrm{c}}\,,&&
\end{eqnarray}
which yields
\begin{eqnarray}
  {\overline{\cal R}}_{\mathrm{c}}=-\frac{1}{2\pi}
  \frac{GmM_{\mathrm{c}}}{\beta R_{\mathrm{CND}}}
  \left(\frac{a}{R_{\mathrm{CND}}}\right)^{\beta}
  \int\limits_{-\pi}^{\pi}\mathrm{d}l\,
  \left(\frac{r}{a}\right)^{\beta},&&
\end{eqnarray}
where $a$ and $e$ are semi-major axis and eccentricity of the stellar
orbit, $r=a\left(1-e\cos{u}\right)$ and $u-e\sin{u}=l$. After an easy
algebra, we obtain
\begin{eqnarray}
  {\overline{\cal R}}_{\mathrm{c}} =
  -\frac{GmM_{\mathrm{c}}}{\beta R_{\mathrm{CND}}}\,
  \left(\frac{a}{R_{\mathrm{CND}}}\right)^{\beta}{\cal J}
  \left(e,\beta\right)\,,&&
\label{bw3}
\end{eqnarray}
where
\begin{eqnarray}
  {\cal J}\left(e,\beta\right)\equiv\frac{1}{\pi}\int\limits_{0}^{\pi}
  \mathrm{d}u\,\left(1-e\cos{u}\right)^{1+\beta}=
  1+\sum_{n\geq 1} a_n e^{2n},&&
\end{eqnarray}
with the coefficients obtained by recurrence
\begin{eqnarray}
  \frac{a_{n+1}}{a_n} = \left[1-\frac{3+\beta}{2(n+1)}\right] \left[1-
  \frac{2+\beta}{2(n+1)}\right]&&
\end{eqnarray}
and an initial value $a_1=\beta\left(1+\beta\right)/4$. For the purpose of
our study, we further set $\beta=1/4$ which corresponds to the equilibrium
model worked out by \citet{Bahcall76}.
%
%
\subsubsection{Circumnuclear disc/ring potential}\label{cnd_kozai}
In the case of perturbation of orbits well below the radius of the CND,
we limit ourselves to account for the quadrupole-tide formulation
(e.g. Kozai 1962; Morbidelli 2002). Octupole or
higher-multipole corrections are possible (e.g. in fact Kozai himself
gives explicit terms up to degree 4; see also Yokoyama et~al. 2003)
but they do not change the conclusions as long as the parameter
$a/R_{\mathrm{CND}}$
is small enough. This is the regime that interests us most.

Given the axial symmetry of the mass distribution of
the perturbing ring, the resulting averaged interaction potential energy
of a particle in the tidal field of the CND (see Kozai 1962)
\begin{align}
  \!\!\!\!\!{\overline{\cal R}}_{\mathrm{CND}}&=-
  \frac{GmM_{\mathrm{CND}}}{16 R_{\mathrm{CND}}}\left(
  \frac{a}{R_{\mathrm{CND}}}\right)^2\Bigl[\left(2+3e^2\right)
  \left(3\cos^2 I-1\right)\nonumber\\
  &\;\;\;\;+15e^2\sin^2 I\cos 2\omega\Bigr]
\label{koz1}
\end{align}
does not depend on longitude of node $\Omega$ but depends on other
orbital elements of the stellar orbit -- eccentricity $e$, inclination $I$ and
argument of pericentre $\omega$.
\begin{figure}
\includegraphics[width=\columnwidth]{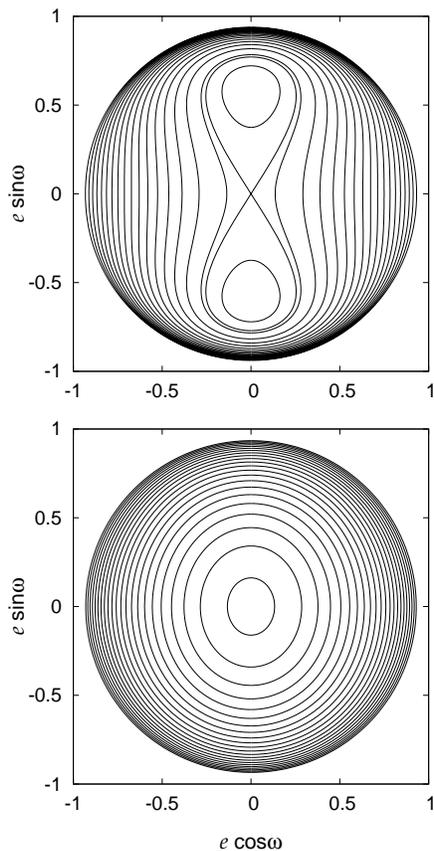}
\caption{Isolines of the conserved potential function
  ${\overline{\cal R}}=C$ from
  equation~(\ref{bwkoz}) for two different values of the
  mass ratio $\mu=M_{\mathrm{c}}/M_{\mathrm{CND}}$:
  0.01 at the top panel, 0.1 at the bottom panel. The
  Kozai integral value is $c=\cos(70^\circ)$, corresponding to $70^\circ$
  inclination circular orbit. The orbit has been given semi-major axis
  $a=0.06\,R_{\mathrm{CND}}$ for sake of definiteness. The
  origin $e=0$ is a stationary point of the problem but in the first
  case it is unstable, while in the second case it becomes stable. The
  thick isoline in the top panel is a separatrix between two different
  regimes of eccentricity and pericentre evolution.}
\label{kozaimodes}
\end{figure}
\begin{figure}
 \includegraphics[width=\columnwidth]{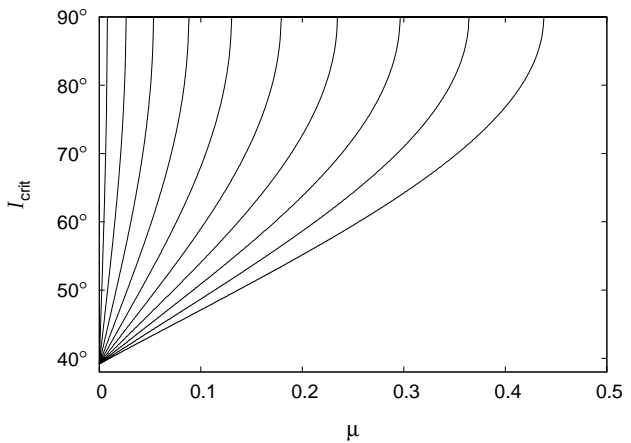}
 \caption{Individual lines show a critical inclination (ordinate) at
  which Kozai resonance onsets for a given value of mass ratio
  $\mu=M_{\mathrm{c}}/M_{\mathrm{CND}}$ (abscissa) for different values of
  orbital semi-major axis $a$ ranging from $0.03\,R_{\mathrm{CND}}$ (left) to
  $0.3\,R_{\mathrm{CND}}$ (right) with the step of
  $0.03\,R_{\mathrm{CND}}$. When $\mu=0$, the critical
  angle is $\approx39.2^{\circ}$ (`the Kozai limit') independently from $a$.}
 \label{ecc}
\end{figure}

As a direct consequence, $c\equiv\sqrt{1-e^2}\cos I$ is the first (`Kozai')
integral of motion, which conveniently allows to eliminate inclination
dependence in ${\overline {\cal R}}_\mathrm{CND}$, depending then on the
eccentricity and argument of pericentre only. Since
${\overline {\cal R}}_\mathrm{CND}$ is a conserved quantity in the secular
(orbit-averaged) problem, the isolines
${\overline {\cal R}}_\mathrm{CND}=C$
provide insights in the fundamental features of the dynamical evolution of
both $e$ and $\omega$. This approach has been used by Kozai to
discover two modes of topology of these isolines:
(i) when $c>\sqrt{3/5}$ the
${\overline {\cal R}}_{\rm CND}=C$ isolines are simple ovals about
origin which is the only fixed point of the problem but (ii) for
$c\leq\sqrt{3/5}$ they become more complicated with a separatrix curve
emerging from the origin and two new fixed points exist at nonzero eccentricity
and pericentre argument values $90^\circ$ and $270^\circ$. The latter
case occurs whenever the initial inclination is larger than $\approx39.2^\circ$,
sometimes called the Kozai limit. The important take-away
message is that the circular orbit is no more a stable solution for
high-inclination orbits in the model of exterior ring/disc perturbation.
Initially circular orbits would be driven over a Kozai timescale
\begin{eqnarray}
  T_{\mathrm{K}}\equiv\frac{M_{\mathrm{\bullet}}}{M_{\mathrm{CND}}}
  \frac{R_{\mathrm{CND}}^3}{a\sqrt{GM_{\bullet}a}}&&
\label{Kozai}
\end{eqnarray}
to a very high eccentricity state. Unavoidable stellar
scattering processes would in a short time destabilize an initially
coherent stream of objects near the centre.
%
%
\subsubsection{Combined perturbation}
We now consider combined effect of the stellar cusp and the CND
potentials on the long-term orbital evolution of the stellar orbit. The
total, orbit-averaged potential
\begin{eqnarray}
  {\overline {\cal R}} = {\overline {\cal R}}_{\rm c}+{\overline
  {\cal R}}_{\rm CND}&&
\label{bwkoz}
\end{eqnarray}
still obeys axial symmetry, being independent on the nodal longitude.
The picture, however, may be modified with respect to the case of
solely ring-like perturbation. Considering the
cusp of the late-type stars whose potential is
approximated with (\ref{bw3}), we find that the two types of topologies
of the ${\overline {\cal R}}=C$ isolines persist (see
Fig.~\ref{kozaimodes}) but the onset of the circular-orbit instability
depends now on two parameters, namely $c$ and $\mu\equiv M_{\rm c}/M_{\rm CND}$.
A nonzero mass of the stellar cusp stabilizes small eccentricity evolution
and the critical angle is pushed to larger values.
For large enough $\mu$, the stability of the circular orbit
is guaranteed for arbitrary value of $c$ and hence orbits of an
arbitrary inclination with respect to the CND symmetry plane. This is
because the effects of the stellar cusp potential make the argument
of pericentre circulate fast enough (significantly faster than the
Kozai timescale), preventing thus secular increase of the eccentricity.
An initially near-circular orbit maintains a very small value of $e$
showing only small-amplitude oscillations. Fig.~\ref{ecc} shows critical
inclination values, for which
the circular orbit becomes necessarily unstable as a function of $\mu$ and
$a/R_{\mathrm{CND}}$ parameters (note the later factorizes out from the analysis
when
$\mu=0$). Importantly, there is a correlation between $\mu$ and
$a/R_{\mathrm{CND}}$ below
which circular orbits of an arbitrary inclination are stable; for
instance, data in Fig.~\ref{ecc} indicate that for $\mu=0.1$ any circular
orbit with $a\lesssim 0.12\,R_{\mathrm{CND}}$ is stable.

In conclusion, we observe that having enough mass in the late-type
stellar cusp may produce strong enough perturbation to maintain small
eccentricity
of an initially near-circular orbit. With that said, we find it reasonable to
make an important simplification within our analytic approach to the system
of two (multiple) stars. Namely, we will further consider the stellar orbits
to be circular during the whole evolution of the system. This prevents
(together with the assumption of well separated orbits with constant
semi-major axes) close encounters of the stars.
In this case only, and under the assumption that there are no orbital
resonances
among the individual stars, the mutual interaction of the stars
may be reasonably considered as a perturbation to the dominating
potential of the SMBH.
As we demonstrate in the next sections, this simple
treatment provides useful insights into the evolution of the young-stream orbits
even if they are generally non-circular.
%
%
\subsection{Orbital evolution of circular orbits}\label{incom}
Having discussed our assumptions about semi-major axes, eccentricity and
pericentre of the
stellar orbits, we may now turn to description of the evolution
of the two remaining orbital elements -- inclination
and nodal longitude.
We start with a model of two interacting stars and later generalize it
to the case of an arbitrary number of stars. The major leap-forward in the
model is that we now take into account also mutual gravitational effects of the
two stars.
On the contrary, note that the orbit-averaged potential energy (\ref{bw3})
of the late-type stellar cusp depends on the semi-major axis and eccentricity
only, and thus does not influence evolution of inclination and node. For that
reason it drops from our analysis in this section.

The interaction potential energy ${\cal R}_{\rm i}(\boldsymbol{r},
\boldsymbol{r}')$ for two
point sources of masses $m$ and $m^{\prime}$
at relative positions $\boldsymbol{r}$ and $\boldsymbol{r}'$ with respect
to the centre reads%
\footnote{Note that equation~(\ref{co1}) provides the interaction energy as it
 appears
 in the equation of relative motion of stars with respect to the centre.
 Henceforth, the perturbation series start with a quadrupole term ($\ell=2$).}
\begin{eqnarray}
  {\cal R}_{\rm i}(\boldsymbol{r},\boldsymbol{r}') =
  -\frac{G m m'}{r}\,\sum_{\ell\geq 2}
  \alpha^\ell P_\ell\left(\cos S\right)\,,&&
\label{co1}
\end{eqnarray}
where $P_\ell(x)$ are Legendre polynomials, $\cos S\equiv\boldsymbol{r}\cdot
\boldsymbol{r}'/r r'$ and $\alpha \equiv r'/r$. The series in
the right-hand side of equation~(\ref{co1}) converge for $r'<r$. Since we are
going to apply (\ref{co1}) to the simplified case of two circular orbits,
we may replace distances $r$ and $r'$ with the corresponding values of
semi-major axis $a$ and $a'$, such that $\alpha=a'/a$ now (note that the
orbit whose parameters are denoted with a prime is thus assumed interior).
The averaging of the
interaction energy over the uniform orbital motion of the stars about the
centre, implying periodic variation of $S$, is readily performed by using the
addition theorem for spherical harmonics. This allows us to decouple unit
direction vectors in the argument of the Legendre polynomial $P_\ell$ and
easily obtain the required average of ${\cal R}_{\rm i}$ over the orbital
periods of the two stars. After a simple algebra we obtain
\begin{eqnarray}
  {\overline {\cal R}}_{\rm i} =
  - \frac{G m m'}{a}\,\Psi\left(\alpha,\boldsymbol{n}\cdot
  \boldsymbol{n}'\right)\,,&&
\label{co2}
\end{eqnarray}
where $\boldsymbol{n}=[\sin I\sin\Omega,-\sin I\cos\Omega,\cos I]^{\rm T}$ and
$\boldsymbol{n}'=[\sin I'\sin\Omega',-\sin I'\cos\Omega',\cos I']^{\rm T}$ are
unit
vectors normal to the mean orbital planes of the two stars, and
\begin{eqnarray}
  \Psi\left(\zeta,x\right) = \sum_{\ell\geq 2} \left[P_\ell\left(0\right)
  \right]^2 \zeta^\ell P_\ell\left(x\right)\,.&&
\label{co3}
\end{eqnarray}
As expected, the potential energy is only a function of: (i) the orbital
semi-major axes through dependence on $a$ and $\alpha$, and (ii) the relative
configuration of the two orbits in space given by the scalar product
$\boldsymbol{n}\cdot\boldsymbol{n}'$. Note also that the series in (\ref{co3})
contain
only even multipoles $\ell$ ($P_\ell(0)=0$ for $\ell$ odd) and that
they converge when $\zeta <1$. However, a special care is needed when
$\zeta$ is very close to unity, thus the two stellar orbits are close to
each other, when hundreds to thousands terms are needed to achieve
sufficient
accuracy. Still, we found it is very easy to set up an efficient computer
algorithm, using recurrent relations between the Legendre polynomials,
which is able to evaluate (\ref{co3}) and its derivatives. In practice,
we select a required accuracy and the code truncates the series by
estimating the remained terms. In fact, since our approach neglects
small eccentricity oscillations of the orbits we are anyway not allowed to
set $\zeta=\alpha=a'/a$ arbitrarily close to unity. Theoretically,
we should require
\begin{eqnarray}
  \alpha < 1 - \left(\frac{m+m'}{3M_{\bullet}}\right)^{1/3},&&
\label{cohill}
\end{eqnarray}
by not letting the stars approach closer than the Hill radius of their
mutual interaction. In the numerical examples we present below, this sets
an upper limit $\alpha < 0.98$.

The formulation given above immediately provides potential energy of the
star-CND interaction. In this case the stellar orbits are always interior
to the CND with symmetry axis suitably chosen
as the unity vector
$\boldsymbol{e}_z$ in the direction of the $z$-axis
of our reference system.
Unlike in Section~\ref{cnd_kozai}, we restrict now to
the case of circular orbit of the star but at the low computer-time
expense we may include all multipole terms till specified accuracy is
achieved.
As a result the orbit-averaged interaction energy
with the exterior stellar orbit is given by
\begin{eqnarray}
  {\overline {\cal R}}_{\rm CND} =
  - \frac{G m M_{\rm CND}}{R_{\mathrm{CND}}}\,\Psi\left(a/R_{\mathrm{CND}},
  \cos I\right)\,,&&
\label{co4}
\end{eqnarray}
and similarly for the interior stellar orbit:
\begin{eqnarray}
  {\overline {\cal R}}'_{\rm CND} =
  - \frac{G m' M_{\rm CND}}{R_{\mathrm{CND}}}\,\Psi\left(a'/R_{\mathrm{CND}},
  \cos I'\right)\,.&&
\label{co5}
\end{eqnarray}
The total orbit-averaged potential energy perturbing motion of the two stars
is then given by superposition of the three terms:
\begin{eqnarray}
  {\overline {\cal R}} = {\overline {\cal R}}_{\rm i} + {\overline {\cal
  R}}_{\rm CND} + {\overline {\cal R}}'_{\rm CND}\;.&&
\label{co6}
\end{eqnarray}
Recalling that semi-major axis values are constant, eccentricity set to zero
and thus argument of pericentre undefined, we are left to study dynamics
of inclination $I$ and $I'$ and longitude of node $\Omega$ and $\Omega'$
values. Lagrange equations provide (see, e.g., Bertotti et~al. 2003)
\begin{align}
  &\!\!\!\!\!\!\!\!
  \frac{\mathrm{d}\cos{I}}{\mathrm{d}t}=-\frac{1}{mna^2}
  \frac{\partial{\overline {\cal R}}}{\partial \Omega}\;,\quad\;\;\;\;\,
  \frac{\mathrm{d}\Omega}{\mathrm{d}t}
  =\frac{1}{mna^2}\frac{\partial
  {\overline {\cal R}}}{\partial \cos I}\;,
\label{co7a}\\
  &\!\!\!\!\!\!\!\!
  \frac{\mathrm{d}\cos{I'}}{\mathrm{d}t} = -
  \frac{1}{m'n'a'^2}\frac{\partial
  {\overline {\cal R}}}{\partial \Omega'}\;,\quad
  \frac{\mathrm{d}\Omega'}{\mathrm{d}t} = \frac{1}{m'n'a'^2}\frac{\partial
  {\overline {\cal R}}}{\partial \cos I'}\;,
\label{co7b}
\end{align}
where $n$ and $n'$ denote mean motion frequencies of the two stars. Note the
particularly simple, quasi-Hamiltonian form of equations~(\ref{co7a}) and
(\ref{co7b}).
They can also be rewritten in a more compact way using the normal vectors
$\boldsymbol{n}$
and $\boldsymbol{n}'$ to the respective orbit, namely
\begin{align}
  &\!\!\!\!\!\!\!\!\!\!\!\!\!\!\!\!\!\!\!\!\!\!\!\!\!\!\!\!\!\!\!\!
  \!\!\!\!\!\!\!\!\!\!\!\!\!\!\!\!\!\!\!\!\!\!\!\!\!\!\!\!\!\!\!\!
  \!\!\!\!\!\!\!\!\!\!\!\!\!\!
  \frac{\mathrm{d} \boldsymbol{n}}{\mathrm{d}t} =
  \boldsymbol{n}\times \frac{\partial}{\partial \boldsymbol{n}}
  \left(\frac{{\overline {\cal R}}}{mna^2}\right)\,,\;
\label{co8a}\\
  &\!\!\!\!\!\!\!\!\!\!\!\!\!\!\!\!\!\!\!\!\!\!\!\!\!\!\!\!\!\!\!\!
  \!\!\!\!\!\!\!\!\!\!\!\!\!\!\!\!\!\!\!\!\!\!\!\!\!\!\!\!\!\!\!\!
  \!\!\!\!\!\!\!\!\!\!\!\!\!\!
  \frac{\mathrm{d} \boldsymbol{n}'}{\mathrm{d}t} =
  \boldsymbol{n}'\times \frac{\partial}{\partial \boldsymbol{n}'}
  \left(\frac{{\overline {\cal R}}}{m'n'a'^2}\right)\,.\;
\label{co8}
\end{align}
Inserting here ${\overline {\cal R}}$ from (\ref{co6}), we finally obtain
\begin{align}
  &\!\!\!\!\!\!\!\!\!\!\!\!\!\!\!\!\!\!\!\!\!\!\!\!\!\!\!\!\!\!\!\!
  \!\!\!\!\!\!\!\!\!\!\!\!\!\!\!\!\!\!\!\!\!\!\!\!\!\!
  \frac{\mathrm{d} \boldsymbol{n}}{\mathrm{d}t} = \omega_{\rm I}
  \left(\boldsymbol{n}\times\boldsymbol{n}'\right)+
  \omega_{\rm CND} \left(\boldsymbol{n}\times\boldsymbol{e}_z\right)\,,
\label{co9a}\\
  &\!\!\!\!\!\!\!\!\!\!\!\!\!\!\!\!\!\!\!\!\!\!\!\!\!\!\!\!\!\!\!\!
  \!\!\!\!\!\!\!\!\!\!\!\!\!\!\!\!\!\!\!\!\!\!\!\!\!\!
  \frac{\mathrm{d} \boldsymbol{n}'}{\mathrm{d}t} = \omega'_{\rm I}
  \left(\boldsymbol{n}'\times\boldsymbol{n}\right)+
  \omega'_{\rm CND} \left(\boldsymbol{n}'\times\boldsymbol{e}_z\right)\,,
\label{co9b}
\end{align}
where
\begin{align}
  &\!\!\!\!\!\!\!\!\!\!\!\!\!\!\!\!\!\!\!\!\!\!\!\!\!\!\!\!\!\!\!\!
  \!\!\!\!\!\!\!\!\!\!\!
  \omega_{\rm I} = -n \phantom{\alpha}\left(\frac{m'}{M_{\bullet}}
  \right)\Psi_x\left(\alpha,\boldsymbol{n}\cdot\boldsymbol{n}'\right)\,,
\label{co10a}\\
  &\!\!\!\!\!\!\!\!\!\!\!\!\!\!\!\!\!\!\!\!\!\!\!\!\!\!\!\!\!\!\!\!
  \!\!\!\!\!\!\!\!\!\!\!
  \omega'_{\rm I} = -n'\alpha \left(\frac{m}{M_{\bullet}}
  \right)\Psi_x
  \left(\alpha,\boldsymbol{n}\cdot\boldsymbol{n}'\right)\,,
\label{co10b}\\
  &\!\!\!\!\!\!\!\!\!\!\!\!\!\!\!\!\!\!\!\!\!\!\!\!\!\!\!\!\!\!\!\!
  \!\!\!\!\!\!\!\!\!\!\!
  \omega_{\rm CND} = -n \left(\frac{M_{\rm CND}}{M_{\bullet}}
  \right)\Psi_x
  \left(a/R_{\mathrm{CND}},n_z\right)\,,
\label{co10c}\\
  &\!\!\!\!\!\!\!\!\!\!\!\!\!\!\!\!\!\!\!\!\!\!\!\!\!\!\!\!\!\!\!\!
  \!\!\!\!\!\!\!\!\!\!\!
  \omega'_{\rm CND} = -n' \left(\frac{M_{\rm CND}}{M_{\bullet}}
  \right)
  \Psi_x\left(a'/R_{\mathrm{CND}},n'_z\right)\,.
\label{co10d}
\end{align}
Note the frequencies in (\ref{co10a}) to (\ref{co10d}) depend on both
$\boldsymbol{n}$
and
$\boldsymbol{n}'$ through their presence in the argument of
\begin{eqnarray}
  \Psi_x(\zeta,x) \equiv \frac{\mathrm{d}}{\mathrm{d}x}\,\Psi(\zeta,x)\,,&&
\label{co11}
\end{eqnarray}
which breaks the apparent simplicity of the system of equations~(\ref{co9a})
and
(\ref{co9b}).

The coupled set of equations~(\ref{co9a}) and (\ref{co9b}) acquires simple
solutions in two limiting cases. First, when $m=m'=0$
(i.e. mutual interaction of stars is neglected) the two equations decouple
and describe simple precession of $\boldsymbol{n}$ and $\boldsymbol{n}'$ about
$\boldsymbol{e}_z$ axis of the inertial frame with frequencies
$-\omega_{\rm CND}
\cos I$ and $-\omega'_{\rm CND}\cos I'$.
The sign minus of these frequencies recalls that the
orbits precess in a retrograde sense when inclinations are less than
$90^\circ$ and vice versa.
Both inclinations $I$ and $I'$
are constant. In the second limit, when $M_{\rm CND}=0$ (i.e. the circumnuclear
torus is removed) the equations~(\ref{co9a}) and (\ref{co9b}) obey a
general integral of total angular momentum conservation
\begin{eqnarray}
  m\, \boldsymbol{n} + m'\alpha^{1/2}\, \boldsymbol{n}'=\boldsymbol{K}\;.&&
\label{co12}
\end{eqnarray}
Both vectors $\boldsymbol{n}$ and
$\boldsymbol{n}'$ then precess about $\boldsymbol{K}$ with
the same frequency
\begin{eqnarray}
  \omega_{\rm p} = \frac{\omega_{\rm I}}{m'\alpha^{1/2}}\frac{m+m'\alpha^{1/2}
  \left(\boldsymbol{n}\cdot\boldsymbol{n}'\right)}{\sqrt{m^2+
  m'^2\alpha+2mm'\alpha^{1/2}
  \left(\boldsymbol{n}\cdot\boldsymbol{n}'\right)}}\;,&&
\label{co13}
\end{eqnarray}
keeping the same mutual configuration. In particular, initially coplanar
orbits (i.e. $\boldsymbol{n}$ and $\boldsymbol{n}'$ parallel) would not evolve,
which is in agreement with intuition.

Unfortunately, we were not able to find analytical solution to the
(\ref{co9a}) and (\ref{co9b}) system except for these two situations
described above. Obviously, it can be always solved using numerical
methods as we shall discuss in Section~\ref{ns}.
\begin{figure}
\includegraphics[width=\columnwidth]{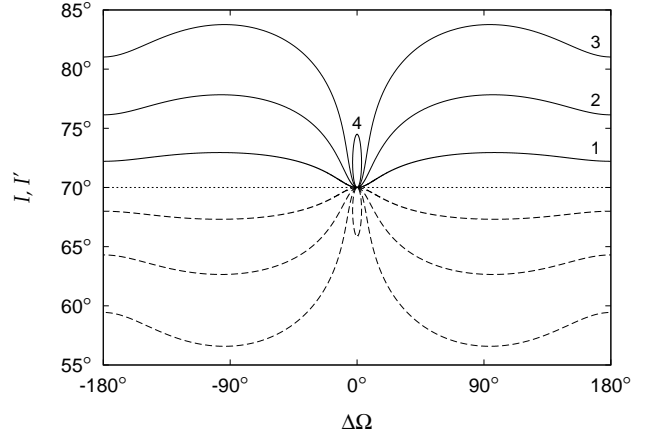}
\caption{Isolines of the ${\overline {\cal R}}=C_2$ integral in the $I$ or
  $I'$ vs.
  $\Delta\Omega$ space. For sake of example we use orbits of two
  equal-mass stars ($m'=m$) with semi-major axes
  $a'=0.04\,R_{\mathrm{CND}}$ and
  $a=0.05\,R_{\mathrm{CND}}$. The mass of the CND is set to
  $M_{\rm CND}=0.3\,M_{\bullet}$.
  The individual lines correspond to
  different values of stellar mass: $m=5\times10^{-7}\,M_{\bullet}$
  (curves~1), $m=2\times10^{-6}\,M_{\bullet}$
  (curves~2), $m=5\times 10^{-6}\,M_{\bullet}$ (curves~3), and
  $m=9\times10^{-6}\,M_{\bullet}$
  (curves~4). Both orbits have been given $70^\circ$ inclination at
  $\Delta\Omega=0^\circ$ (i.e. initially coplanar and inclined orbits).
  Solid lines show inclination $I'$ of the inner orbit, `the mirror-imaged'
  dashed lines describe inclination $I$ of the outer orbit.}
\label{int}
\end{figure}
\begin{figure*}
  \includegraphics[width=\textwidth]{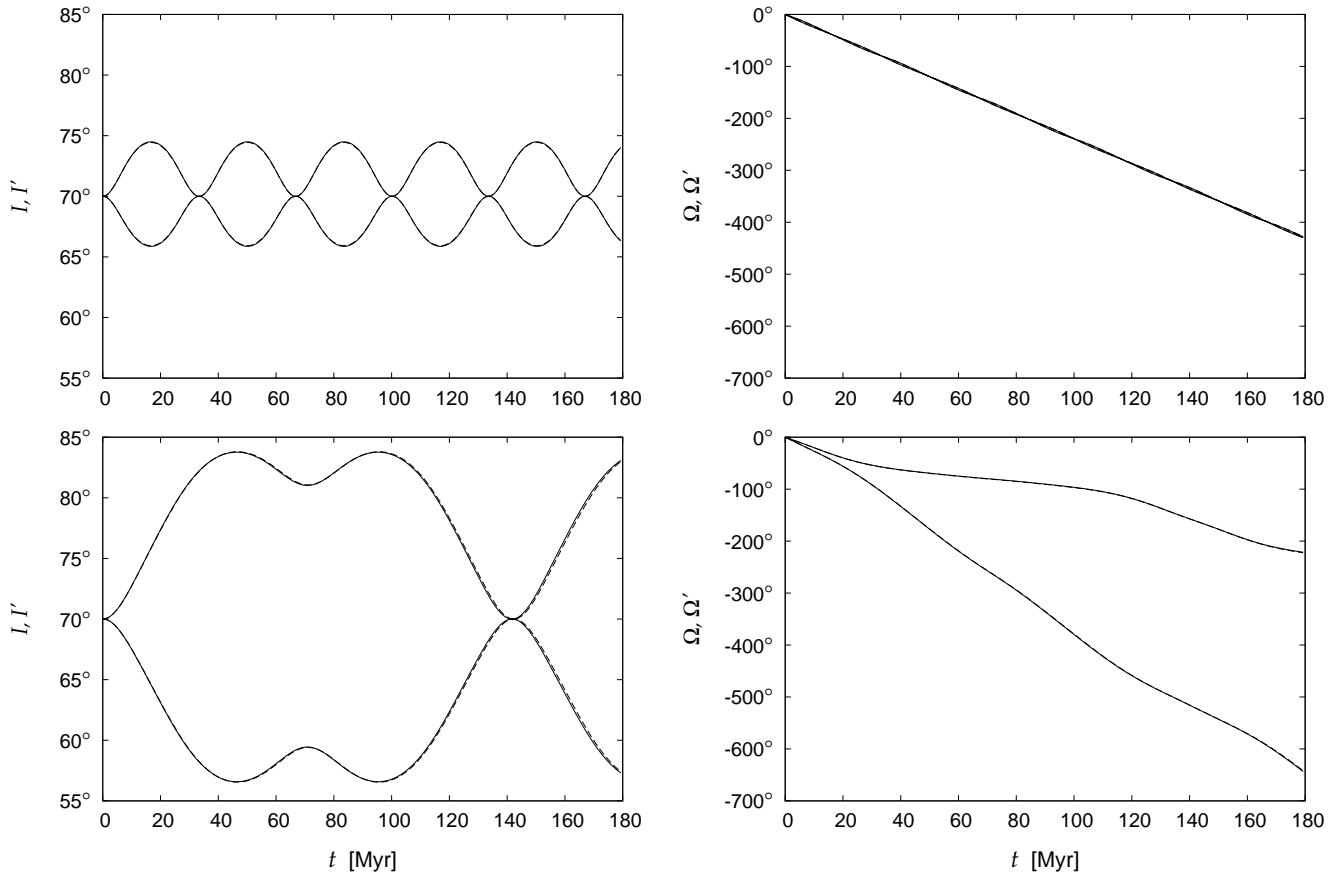}
  \caption{Evolution of the system of two stars in the compound potential
  of the central SMBH, spherical stellar cusp and axisymmetric
  CND. Solid lines represent solution of two-body equations~(\ref{co9a}) and
  (\ref{co9b}), while the dashed lines show result of the direct numerical
  integration of the equations of motion. In each panel, upper and lower lines
  correspond to the inner and outer star, respectively. Common parameters
  for both examples are the same as in Fig.~\ref{int}; in the upper panels,
  we set $m=m^{\prime}=9\times10^{-6}\,M_{\bullet}$,
  while in the lower ones $m=m^{\prime}=5\times10^{-6}\,M_{\bullet}$.}
  \label{two_stars}
\end{figure*}
%
%
\subsubsection{Integrals of motion}\label{im}
In general, equations~(\ref{co9a}) and (\ref{co9b}) have only two first
integrals. Our assumptions about the circumnuclear torus mass
distribution still
provide a symmetry vector $\boldsymbol{e}_z$. Thus, while the total angular
momentum $\boldsymbol{K}$ is no more conserved now, its projection onto
$\boldsymbol{e}_z$ is still an integral of motion
\begin{eqnarray}
  m\cos I + m'\alpha^{1/2}\cos I'=C_1 =K_z\;.&&
\label{co14}
\end{eqnarray}
Because $m$, $m'$ and $\alpha$ are constant, equation~(\ref{co14}) provides
a direct constraint of how the two inclinations $I$ and $I'$ evolve.
In particular, one can be expressed as a function of the other.
\begin{figure*}
  \includegraphics[width=\textwidth]{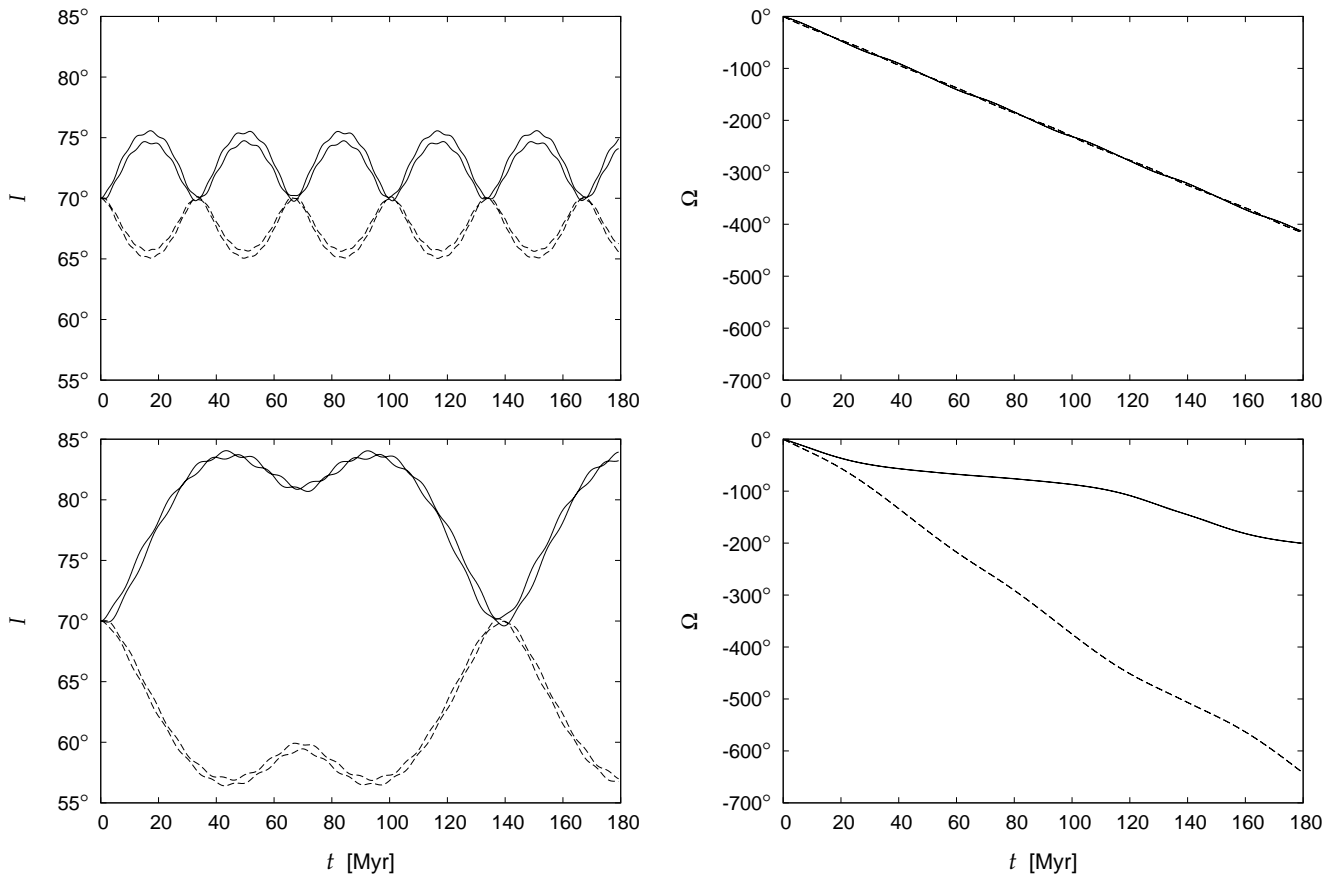}
  \caption{Evolution of the system of four stars
  in the compound potential
  of the central SMBH, spherical stellar cusp and axisymmetric
  CND. The stellar orbits form two couples. In both of them, the orbits
  have similar semi-major axes in order to mimic the system shown in
  Fig.~\ref{two_stars}. In each panel, upper and lower lines
  correspond to the inner and outer couple, respectively. The individual
  semi-major axes are for both examples set to
  $a_1=0.0373\,R_{\mathrm{CND}}$, $a_2=0.0408\,R_{\mathrm{CND}}$,
  $a_3=0.0478\,R_{\mathrm{CND}}$, $a_4=0.0511\,R_{\mathrm{CND}}$.
  The other common parameters
  for both examples are the same as in Fig.~\ref{int}; in the upper panels,
  we set $m_1=m_2=m_3=m_4=4.5\times10^{-6}\,M_{\bullet}$,
  while in the lower ones $m_1=m_2=m_3=m_4=2.5\times10^{-6}\,M_{\bullet}$.}
  \label{regular}
\end{figure*}

The quasi-Hamiltonian form of equations~(\ref{co7a}) and (\ref{co7b})
readily results in a second integral of motion
\begin{eqnarray}
  {\overline {\cal R}}
  \left(\cos I,\cos I',\boldsymbol{n}\cdot\boldsymbol{n}'\right)
  = C_2\;.&&
\label{co15}
\end{eqnarray}
The list of arguments in ${\overline {\cal R}}$, as explicitly
provided above, reminds that it actually depends on: (i) the inclination
values $I$ and $I'$, and (ii) the difference $\Delta \Omega=\Omega-\Omega'$
of the nodal longitudes of the two interacting orbits. Using (\ref{co14}),
the conservation of ${\overline {\cal R}}$ thus provides a constraint
between the evolution of $I$ and $\Delta \Omega$ (say). While not giving
a solution of the problem, the constraint due to combination of first
integrals (\ref{co14}) and (\ref{co15}) can still provide useful insights.

Fig.~\ref{int} illustrates how the first integrals help understanding
several features of the orbital evolution for two interacting stars at
distances $a'=0.04\,R_{\mathrm{CND}}$ and $a=0.05\,R_{\mathrm{CND}}$.
For sake of simplicity we also assume their
mass is equal, hence $m'=m$, and the CND has
been given mass $M_{\rm CND}=0.3\,M_{\bullet}$. Data in this figure show
constrained
evolution of orbital inclinations $I'$ (solid lines) and $I$ (dashed lines)
as a function of nodal difference $\Delta \Omega$. The two orbits were
assumed to be initially coplanar ($\Delta\Omega=0^\circ$) with an
inclination of $I'=I=70^\circ$. A set of curves correspond to different
values of stellar masses, from small (1) to larger values (4), which basically
means increasing strength of their mutual gravitational interaction.

First, conservation of the $\boldsymbol{e}_z$-projected orbital angular
momentum,
as given by equation~(\ref{co14}), requires that increase in $I'$ is compensated
by decrease of $I$. This results in a near-mirror-imaged evolution of the two
inclinations. Using the first equation of (\ref{co7a}), one finds
\begin{eqnarray}
  \frac{\mathrm{d}I}{\mathrm{d}t} =
  \frac{n}{\sin I}\frac{m'}{M_{\bullet}}\sin\left(\Omega-
  \Omega'\right) \Psi_x
  \left(\alpha,\boldsymbol{n}\cdot\boldsymbol{n}'\right)\,,&&
\label{co16}
\end{eqnarray}
which straightforwardly implies that the outer stellar orbit is initially
torqued to decrease its inclination while the inner orbit increases its
inclination. This is because initially
$\boldsymbol{n}\cdot\boldsymbol{n}'\approx1$, and
$\Psi_x(\alpha,1)$ is positive, and at the same time, precession of the
nodes is dominated by interaction with the CND which makes the outward orbit
node to drift faster (and hence $\Omega-\Omega'$ is negative).

Second, Fig.~\ref{int} indicates there is important change in topology
of the isolines ${\overline {\cal R}}=C_2$ as the stellar masses overpass
some critical value (about $8.5\times10^{-6}\,M_{\bullet}$ in our example).
For low-mass
stars their mutual gravitational interaction is weak letting the effects
of the CND dominate (curve 1). The orbits regularly precess with different
frequency, given their different distance from the centre, and thus
$\Delta\Omega$ acquires all values between $-180^\circ$ and $180^\circ$.
The mutual stellar interaction produces only small inclination oscillation.
As the stellar masses increase (curves 2 and 3) the inclination perturbation
becomes larger. For super-critical values of $m$ (curve 4) the isolines
of constant ${\overline {\cal R}}$ become only small loops about the origin.
This means that $\Delta\Omega$ is bound to oscillate in a small interval
near origin and inclination perturbation becomes strongly damped. Put in
words, the gravitational coupling between the stars became strong
enough to
tightly couple the two orbits together. Note that they still collectively
precess in space due to the influence of the CND.
%
%
\subsubsection{Numerical solutions} \label{ns}
In order to solve equations~(\ref{co9a})
and (\ref{co9b}) numerically, we adopt a
simple adaptive step-size 4.5th-order Runge-Kutta algorithm.
Let us mention that our implementation of this algorithm
conserves the value of both integrals of motion $C_1$ and $C_2$ with relative
accuracy better than $10^{-6}$.

Two sample solutions are shown in Fig.~\ref{two_stars}.
The upper panels represent evolution of two orbits with coupled precession
which corresponds to the curve~4 in Fig.~\ref{int}, while
in the bottom panels we consider the case of lower-mass stars, whose
orbits precess independently. This later mode corresponds to the curve 3 in
Fig.~\ref{int}. Beside the solution of the equations for mean orbital elements,
we also show results of a full-fledged numerical integration of the
particular configuration in the space of classical positions and momenta
($\boldsymbol{r}, \boldsymbol{r}^\prime; \boldsymbol{p},
\boldsymbol{p}^\prime$). Both solutions
are nearly identical, which confirms validity of the
secular perturbation theory used in this paper.
\begin{figure*}
  \includegraphics[width=\textwidth]{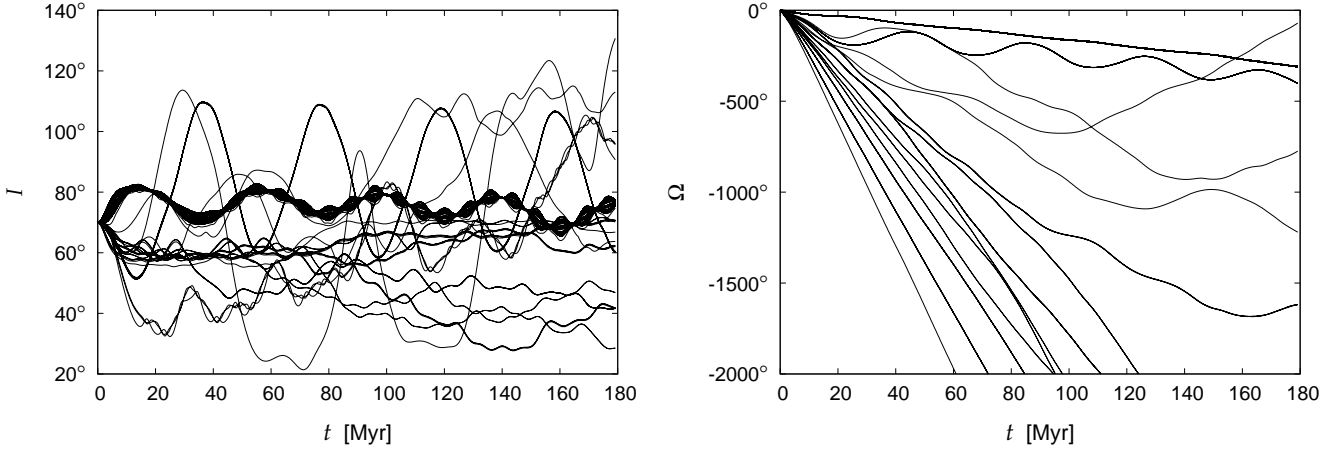}
  \caption{Evolution of the initially thin stellar disc of 100 stars
  in the compound potential
  of the central SMBH, spherical stellar cusp and axisymmetric
  CND. The values of orbital semi-major axes $a_k$ in the disc
  range from
  $0.02\,R_{\mathrm{CND}}$ to $0.2\,R_{\mathrm{CND}}$ and their
  distribution obeys
  $\mathrm{d}N\propto a^{-1}\mathrm{d}a$. The stellar masses are all equal
  with $m=5\times10^{-6}\,M_{\bullet}$ while the mass of the CND is set to
  $M_{\mathrm{CND}}=0.3\,M_{\bullet}$. Initial
  inclination $I_0$ of all the orbits with respect
  to the CND equals $70^{\circ}$.}
  \label{chaos}
\end{figure*}

For sake of further discussion we find it useful to comment in
a little more detail on the case of two, nearly independently precessing
orbits (bottom panels on Fig.~\ref{two_stars}). In this case, the precession
frequencies of the outer and inner star orbits are given by $\omega_{\rm
CND}$ and $\omega'_{\rm CND}$ in equations~(\ref{co10c}) and (\ref{co10d}).
When truncated to the quadrupole ($\ell=2$) level, sufficient for the
small value of $a/R_{\mathrm{CND}}$, one has for the
outer star orbit
\begin{eqnarray}
  \frac{\mathrm{d}\Omega}{\mathrm{d}t}\simeq
  -\frac{3}{4}\frac{\cos{I}}{T_{\mathrm{K}}}\;,&&
\label{pre1}
\end{eqnarray}
where $T_{\mathrm{K}}$ is given by (\ref{Kozai}).
A similar formula holds for the inner star orbit
denoted with primed variables. As seen in Fig.~\ref{int}, and understood
from the analysis of integrals of motion in Section~\ref{im}, period
of the evolution of the system of the two orbits is given implicitely
by the difference of their precession rate: $\Omega(T_{\mathrm{char}}) -
\Omega^{\prime}(T_{\mathrm{char}}) = 2\pi$. The secular rate of nodal
precession in (\ref{pre1}) is not constant because the mutual gravitational
interaction of the stars makes their orbital inclinations oscillate. However,
in the zero approximation we may replace them with their initial values,
$I = I^\prime = I_0$ which gives an order of magnitude estimate
\begin{eqnarray}
  T_{\mathrm{char}} \simeq \frac{8\pi}{3\,\cos I_0} \left[\frac{1}{
  T_{\mathrm{K}}}-\frac{1}{T_{\mathrm{K}}^{\prime}}\right]^{-1}.&&
\label{char}
\end{eqnarray}
For the solution shown in the lower panels of Fig.~\ref{two_stars},
formula~(\ref{char}) gives $T_{\mathrm{char}}\approx460$~Myr,
in a reasonable agreement with the observed period of $\approx140~\myr$.
When the orbital evolution is known (being integrated numerically), more
accurate estimate can be obtained considering mean values of the inclinations
\begin{eqnarray}
  T_{\mathrm{char}} \simeq \frac{8\pi}{3}
  \left[\frac{\cos {\overline I}}{T_{\mathrm{K}}}-
  \frac{\cos {\overline I}'}{T_{\mathrm{K}}^{\prime}}\right]^{-1}.&&
\label{char2}
\end{eqnarray}
For the case of the solution of the lower panel of Fig.~\ref{two_stars}, with
${\overline I}\approx60^\circ$ and ${\overline I}'\approx80^\circ$,
formula~(\ref{char2}) gives $T_{\mathrm{char}}\approx120~\myr$.
%
%
\subsubsection{Generalization for N interacting stars}
\label{sec:Nstars}
The previous formulation straightforwardly generalizes to the case of
$N$ stars orbiting the centre on circular orbits with semi-major axes $a_k$
($k=1,\ldots,N$). This is because the potential energies of all
pairwise interactions built the total
\begin{eqnarray}
  {\overline {\cal R}}_{\rm i} = - \frac{1}{2} \sum_{k\neq l}
  \frac{G m_k m_l}{a_{kl}}\,\Psi\left(\alpha_{kl},\boldsymbol{n}_k\cdot
  \boldsymbol{n}_l\right)\,,&&
\label{co17}
\end{eqnarray}
where $m_k$ is the mass of the $k$-th star, $a_{kl} = {\rm min}(a_k,a_l)$,
$\alpha_{kl}={\rm min}(a_k,a_l)/{\rm max}(a_k,a_l)$ and $\boldsymbol{n}_k$
is the normal vector to the orbital plane of the $k$-th star. Similarly,
interaction with the CND is simply given by
\begin{eqnarray}
 {\overline {\cal R}}_{\rm CND} = -\sum_{k}
  \frac{G m_k M_{\rm CND}}{a_{k}}\,\Psi
 \left(a_{k}/R_{\mathrm{CND}},\boldsymbol{n}_k\cdot
  \boldsymbol{e}_z\right)\,.&&
\label{co18}
\end{eqnarray}
The total potential energy of perturbing interactions is
\begin{eqnarray}
  {\overline {\cal R}} = {\overline {\cal R}}_{\rm i} +
  {\overline {\cal R}}_{\rm CND}\;,&&
\label{co19}
\end{eqnarray}
and the equations of orbital evolution now read
\begin{eqnarray}
  \frac{\mathrm{d}\boldsymbol{n}_k}{\mathrm{d}t} =
  \boldsymbol{n}_k\times \frac{\partial}{\partial \boldsymbol{n}_k}
  \left(\frac{{\overline {\cal R}}}{m_k n_k a_k^2}\right)\,,&&
\label{co20}
\end{eqnarray}
for $k=1,\ldots,N$ ($n_k$ is the frequency of the unperturbed mean motion
of the $k$-th star
about the centre). Their first integrals then can be written as
\begin{eqnarray}
  \sum_k m_k n_k a_k^2 \left(\boldsymbol{n}_k\cdot\boldsymbol{e}_z\right)
   = C_1 = K_z&&
\label{co21}
\end{eqnarray}
and
\begin{eqnarray}
  {\overline {\cal R}} = C_2\;.&&
\label{co22}
\end{eqnarray}

Due to mutual interaction of multiple stars, solutions of
equations~(\ref{co20}) represent, in general, an intricate orbital
evolution, whose
course is hardly predictable as it strongly depends upon the initial
setup. Our numerical
experiments show, however, that it is still possible to identify several
qualitative features which remain widely valid. For instance,
a group of orbits with small
separations may orbitaly couple together and effectively act as a single
orbit in interaction with the rest of the stellar system.

This is demonstrated in Fig.~\ref{regular} which shows two
sample solutions of equations~(\ref{co20})
for a system of two such groups. For sake of
clarity, each group consists only of two orbits. Individual semi-major
axes are, for both solutions, set to $a_1=0.0373\,R_{\mathrm{CND}}$,
$a_2=0.0408\,R_{\mathrm{CND}}$,
$a_3=0.0478\,R_{\mathrm{CND}}$, $a_4=0.0511\,R_{\mathrm{CND}}$ in order
to mimic the two-orbits
models from Fig.~\ref{two_stars}. For the same reason,
all the individual masses are considered equal, $m_1=m_2=m_3=m_4$, and set to
$2.5\times10^{-6}\,M_{\bullet}$ in the lower panels, while for the
upper panels we assume
$4.5\times10^{-6}\,M_{\bullet}$. The other parameters remain
identical to the case of
the two-orbits models.
As we can see (cf. Figs~\ref{regular} and \ref{two_stars}),
the dynamical impact of each coupled pair of orbits
upon the rest of the stellar system
is equivalent to the effect of the corresponding single orbit if both the
total mass and semi-major axis of the pair are appropriate.
The individual orbits within each pair then naturally oscillate about the
single-orbit solution according to their mutual interaction.
This conclusion remains valid
even in more complicated systems
as we shall show in the next section.
%
%
\section{Application to the young stellar system in the Sgr A* region}
\label{sec:sgra}
In order to
illustrate the complexity of solutions of equations~(\ref{co20}),
let us now analyze the evolution of a system
which contains an initially thin stellar disc with a
distribution of semi-major
axes of the orbits
$\mathrm{d}N\propto a^{-1}\mathrm{d}a$.
As we can see in Fig.~\ref{chaos},
the oscillations
of the orbital
inclinations no longer have the simple patterns which we observed for the
models discussed in the previous paragraphs. On the other hand, we still can
identify a well defined group of orbits which
coherently change their orientation with respect to the CND. These orbits thus
form a rather thin disc during the whole monitored period of time.
It turns out that they represent
the innermost parts of the initial disc where the separations of the
neighbouring orbits are small enough for their mutual interaction to couple
them together.

The configuration considered in Fig.~\ref{chaos}
roughly matches the main qualitative features of an astrophysical system
which
is observed in the centre of the
Milky Way. It contains a group of early-type stars orbiting
the SMBH on nearly Keplerian orbits. Observations suggest that
about one half of them form
a coherently rotating disc-like structure with estimated surface density
profile $\Sigma\propto R^{-2}$ \citep{Paumard06, Lu09, Bartko09} which
implies the above considered distribution of semi-major axes.
The rest of the early-type stars then appear to be on randomly oriented
orbits.
Both the origin and observed configuration of these
stars represent rather puzzling questions. Due to strong tidal field of the
SMBH, it is impossible for a star to be formed in this region by any
standard star formation mechanism. On the other hand, as the observed stars
are
assumed to be young, no usual transport mechanism is efficient enough
to bring them from farther regions,
where their formation would be
less intricate, within their estimated lifetime.
One of the most promising scenarios of their origin thus considers
formation in situ, via
fragmentation of a self-gravitating gaseous disc \citep{Levin03}.
However, since this process naturally forms stars in a single disc-like
structure, it does not explain the origin of the stars
observed outside the disc. Hence,
in order to justify the in-disc scenario of the formation of the early-type
stars in the Galactic Centre,
some mechanism that may have dragged some of them out from
the parent stellar disc plane is needed.

In our previous paper \citep{Haas11}, we have discussed a possibility that
all the early-type stars had been born in a single disc
which has been, subsequently, partially disrupted by the gravity of the
CND. We have considered the same configuration
of the sources of the gravitational field as
in the current paper and followed the evolution of the disc by means of
direct $N$-body integration. We have observed
coherent evolution of the inner dense part of the disc which exhibited
a tendency to increase its inclination with respect to the CND.
On the other hand, most of the
orbits of the outer parts of the initially coherently rotating disc precessed
independently due to the influence of the CND and, consequently,
became detached from the
parent structure. This behaviour is in accord with
the analysis presented in the current paper.

Furthermore, we can now calculate
the order of magnitude characteristic
time-scale for the `canonical' model of \citet{Haas11} whose system parameters
read:
$M_{\bullet}=4\times10^6\,M_{\odot}$,
$R_{\mathrm{CND}}=1.8~\mathrm{pc}$,
$M_{\mathrm{CND}}=0.3~M_{\bullet}$,
$M_{\mathrm{c}}=0.03~M_{\bullet}$, and
$I_0=70^{\circ}$.
In order to determine the rough time estimate, we use formula (\ref{char}).
As this formula has been derived for
a system of two stars, we
replace the stellar disc with two characteristic particles at certain
radii $a'$, $a$
in the sense of Section~\ref{sec:Nstars}. For this
purpose, let us divide the stars in the disc into two groups according to
their initial distance from the centre and define $a'$ and $a$ as the radii
of the orbits of the median stars in the inner and outer group, i.e.
$a^{\prime}=0.06~\mathrm{pc}$ and
$a=0.23~\mathrm{pc}$. Inserting these values into formula (\ref{char}),
we obtain
$T_{\mathrm{char}}\approx37~\mathrm{Myr}$ for the `canonical' model.
This value is in order of magnitude
agreement with the estimated age of the early-type stars,
$\approx6~\mathrm{Myr}$
\citep{Paumard06}, since the core of the disc reaches its maximal
inclination with respect to the CND
already after a fraction of period
$T_{\mathrm{char}}$ as can be seen in Figs.~\ref{regular} and \ref{chaos}.

Let us emphasize that the results reported in our previous paper
\citep{Haas11} have been acquired by means of full-fledged numerical
integration of equations of motion. As a consequence,
both the eccentricities
and semi-major axes of the individual stellar orbits in the disc
have been naturally undergoing
a significant evolution due to two-body relaxation of the disc.
Moreover, our prior numerical computations have also confirmed
that results similar to those obtained
for the `canonical' model are valid for a wide set of models with
different system parameters, including
the case with zero mass, $M_\mathrm{c}$, of the spherical cusp of the
late-type stars. In the later case, the orbital eccentricities
and inclinations within the stellar disc are subject to high-amplitude
Kozai oscillations.
In conclusion, it appears that
the inner part of the disc
may evolve coherently for a certain period of time
even when we cannot assume neither zero nor
small eccentricity of the stellar orbits. We, therefore, suggest that
also some of the
key qualitative predictions of the semi-analytic theory developed in
the current paper under
the simplifying assumption of circular orbits may be carefully applied to
more general, non-circular systems.

Finally, let us mention that, in addition to the core of the disc,
less significant
groups of orbits with coherent secular evolution may
exist even in the outer parts
of the disc if their separations are small enough. Our semi-analytic approach
thus admits possible existence of secondary disc-like structures
in the observed young stellar system which has indeed been discussed by
several authors \citep{Genzel03, Paumard06, Bartko09}.
%
%
\section{Conclusions}
\label{sec:conclusions}
We have investigated the secular orbital evolution of a system of
$N$ mutually
interacting stars on nearly-circular orbits around the dominating central mass,
considering the perturbative gravitational influence of
a distant axisymmetric source and an extended spherical potential.
Given the
spherical potential is strong enough, we have shown that the secular evolution
of initially circular orbits reduces to the evolution of inclinations
and nodal longitudes. The spherical potential itself can then be
factorized out from the outcoming momentum equations.
Since we have not been able, in a general case, to solve the derived equations
analytically, we have
set up an integrator for their efficient numerical solution. The acquired
results have then been, in order to confirm their validity,
compared with the corresponding full-fledged
numerical integrations in the
space of classical positions and momenta, showing a remarkable agreement.

Some fundamental features of the possible solutions of the new equations
can be understood by an analysis of the integrals of motion.
In the case of
the simplest possible system of two stars interacting in the considered
perturbed
potential, we have identified
two qualitatively different modes of its secular evolution.
If the interaction of the stars is
weak (yet still non-zero),
the secular evolution of their orbits is dominated
by an independent nodal precession. Difference of the individual
precession rates then determines the period of oscillations of the
orbital inclinations.
On the other hand, when
the gravitational interaction of the stars is sufficiently strong (depending
on their mass and the
radii of their orbits), the secular evolution of the
orbits becomes dynamically coupled and, consequently, they
precess coherently around
the symmetry axis of the gravitational potential.
Oscillations of the orbital inclinations are, in this case,
considerably damped.

We have further confirmed, by means of numerical integration of the derived
momentum
equations, that the coupling of strongly interacting orbits
is a generic process that
may occur even in more complex $N$-body systems. In particular, a subset of
stars with strong mutual interaction evolves coherently and, as a result,
its dynamical impact upon the rest of the $N$-body system is similar
to the effect of a single particle of suitable mass and orbital radius.

As an example, we have investigated evolution of a disc-like
structure that roughly models the young stellar system which is
observed in the Galactic Centre. It has turned out that
the semi-analytic work presented in this paper provides a
physical background for understanding of the processes
discovered, by means of
full $N$-body integration, in \cite{Haas11}. Namely, coupling of the strongly
interacting stars from the inner parts of the disc leads to their coherent
orbital evolution, which allows us to observe a disc-like structure even after
several million years of dynamical evolution in the tidal field of the CND.
Orientation of this surviving disc then inevitably changes
towards higher inclination
with respect to the CND, which is in accord
with the observations. On the other
hand, stellar orbits from the outer parts of the disc evolve individually, being
gradually
stripped out
from the parent thin disc structure. Hence, it appears possible
for the
puzzle of the origin of the young stars in the Galactic Centre to be
solved by the
hypothesis of their formation via fragmentation of a single gaseous disc, as
already suggested in \v{S}ubr, Schovancov\'{a} \& Kroupa (2009) and
\cite{Haas11}.

Note that,
beside the physical
explanation of the processes observed in our previous work, the current approach
would
be, due to its low numerical demands,
useful for extensive scanning of the parameter space in order to
confront our model with the observations more thoroughly.
This is going to be a subject of our
future
work when
more accurate observational data will be available.

Finally, let us mention that our semi-analytic model has been developed
under
several simplifying assumptions. Most importantly, the torus CND has been
considered stationary and the cusp of the late-type stars spherically
symmetric. If any of these assumptions were violated, the results
might be more or less affected. For example, a possible anisotropy
of the cusp of the late-type stars
due to chance alignment of some of its stars would break its spherical
symmetry. In that case, the resulting gravitational torques
might have a considerable impact on the dynamical evolution of the stellar disc
as shown by \citet{Kocsis11}.
However, since the
current observational data do not show evidence for such violations,
we may consider our model physically plausible. Moreover,
the currently
available data do suggest roughly perpendicular mutual orientation of the
CND and the stellar disc, which is in accord with the predictions of
both our numerical and semi-analytic model.
We consider this as a supporting argument
for our findigs.
%
%
\section*{Acknowledgments}
We thank the anonymous referee for useful comments.
This work was supported by the Czech Science Foundation
via grants GACR-205/09/H033, GACR-205/07/0052 and GACR-202/09/0772,
from the Research Program MSM0021620860
of the Czech Ministry of Education, and also from project
367611 of the Grant Agency of Charles University in Prague.
The calculations were performed on
the computational cluster Tiger at the Astronomical Institute of Charles
University in Prague
(http://sirrah.troja.mff.cuni.cz/tiger).
%
%

%
\bsp
\label{lastpage}

\begin{thebibliography}{99}
  \bibitem[\protect\citeauthoryear{Bahcall \& Wolf}{1976}]{Bahcall76}
    Bahcall J. N., Wolf R. A., 1976, ApJ, 209, 214
  \bibitem[\protect\citeauthoryear{Bartko et al.}{2009}]{Bartko09}
    Bartko H. et al., 2009, ApJ, 697, 1741
  \bibitem[\protect\citeauthoryear{Bartko et al.}{2010}]{Bartko10}
    Bartko H. et al., 2010, ApJ, 708, 834
  \bibitem[\protect\citeauthoryear{Bertotti et al.}{2003}]{Bertotti03}
    Bertotti B., Farinella P., Vokrouhlick\'{y} D., 2003, Physics of the Solar
    System, Kluwer Academic Publishers, Dordrecht, The Netherlands
  \bibitem[\protect\citeauthoryear{Christopher et al.}{2005}]{Christopher05}
    Christopher M. H., Scoville N. Z., Stolovy S. R., Yun M. S., 2005,
    ApJ, 622, 346
  \bibitem[\protect\citeauthoryear{Do et al.}{2009}]{Dot09}
    Do T., Ghez A. M., Morris M. R., Lu J. R., Matthews K., Yelda S., Larkin J.,
    2009, ApJ, 703, 1323
  \bibitem[\protect\citeauthoryear{Eisenhauer et al.}{2005}]{Eisenhauer05}
    Eisenhauer F. et al., 2005, ApJ, 628, 246
  \bibitem[\protect\citeauthoryear{Genzel et al.}{2003}]{Genzel03}
    Genzel R. et al., 2003, ApJ, 594, 812
  \bibitem[\protect\citeauthoryear{Ghez et al.}{2003}]{Ghez03}
    Ghez A. M. et al., 2003, ApJ, 586, L127
  \bibitem[\protect\citeauthoryear{Ghez et al.}{2005}]{Ghez05}
    Ghez A. M., Salim S., Hornstein S. D., Tanner A., Lu J. R.,
    Morris M., Becklin E. E., Duch\^{e}ne G., 2005, ApJ, 620, 744
  \bibitem[\protect\citeauthoryear{Gillessen et al.}{2009a}]{Gillessen09a}
    Gillessen S., Eisenhauer F., Trippe S., Alexander T., Genzel R., Martins F.,
    Ott T., 2009a, ApJ, 692, 1075
  \bibitem[\protect\citeauthoryear{Gillessen et al.}{2009b}]{Gillessen09b}
    Gillessen S., Eisenhauer F., Fritz T. K., Bartko H., Dodds-Eden K.,
    Pfuhl O., Ott T., Genzel R., 2009b, ApJ, 707, L114
  \bibitem[\protect\citeauthoryear{Haas et al.}{2011}]{Haas11}
    Haas J., \v{S}ubr L., Kroupa P., 2011, MNRAS, 412, 1905
  \bibitem[\protect\citeauthoryear{Ivanov et al.}{2005}]{Ivanov05}
    Ivanov P. B., Polnarev A. G., Saha P., 2005, MNRAS, 358, 1361
  \bibitem[\protect\citeauthoryear{Karas \& \v{S}ubr}{2007}]{Karas07}
    Karas V., \v{S}ubr L., 2007, A\&A, 470, 11
  \bibitem[\protect\citeauthoryear{Kocsis \& Tremaine}{2011}]{Kocsis11}
    Kocsis B., Tremaine S., 2011, MNRAS, 412, 187
  \bibitem[\protect\citeauthoryear{Kozai}{1962}]{Kozai62}
    Kozai Y., 1962, AJ, 67, 591
  \bibitem[\protect\citeauthoryear{Levin \& Beloborodov}{2003}]{Levin03}
    Levin Y., Beloborodov A. M., 2003, ApJ, 590, L33
  \bibitem[\protect\citeauthoryear{Lidov}{1962}]{Lidov62}
    Lidov M. L., 1962, Planetary and Space Sci., 9, 719
  \bibitem[\protect\citeauthoryear{Lu et al.}{2009}]{Lu09}
    Lu J. R., Ghez A. M., Hornstein S. D., Morris M. R., Becklin E. E.,
    Matthews K., 2009, ApJ, 690, 1463
  \bibitem[\protect\citeauthoryear{Morbidelli}{2002}]{Morbidelli02}
    Morbidelli A., 2002, Modern Celestial Mechanics, Taylor \& Francis,
    London and New York
  \bibitem[\protect\citeauthoryear{Paumard et al.}{2006}]{Paumard06}
    Paumard T. et al., 2006, ApJ, 643, 1011
  \bibitem[\protect\citeauthoryear{Sch\"{o}del et al.}{2007}]{Schoedel07}
    Sch\"{o}del R. et al., 2007, A\&A, 469, 125
  \bibitem[\protect\citeauthoryear{\v{S}ubr et al.}{2009}]{Subr09}
    \v{S}ubr L., Schovancov\'{a} J., Kroupa P., 2009, A\&A, 496, 695
  \bibitem[\protect\citeauthoryear{Yelda et al.}{2010}]{Yelda10}
    Yelda S., Ghez A. M., Lu J. R., Do T., Clarkson W., Matthews K., 2010,
    in M. Morris, D. Q. Wang, F. Yuan, eds, Proc. Conf., The Galactic Center:
    A Window on the Nuclear Environment of Disk Galaxies. Astron. Soc. Pac.,
    San Francisco
  \bibitem[\protect\citeauthoryear{Yokoyama et al.}{2003}]{Yokoyama03}
    Yokoyama T., Santos M. T., Cardin G., Winter O. C., 2003, A\&A, 401, 763
\end{thebibliography}
\end{document}